# Laminar and turbulence forced heat transfer convection correlations inside tubes. A review

By:

Gubran A.Q. Abdulrahman[a,b], Sultan M. Alharbi[a]

[a] Mechanical Engineering Department, King Fahd University of Petroleum & Minerals, Dhahran 31261, Saudi Arabia

[b] Interdisciplinary Research Center for Hydrogen Technology and Carbon Management (IRCHTCM), King Fahd University of Petroleum and Minerals, Dhahran 31261, Saudi Arabia

**Abstract**

This work proposes an extensive review of laminar and turbulent forced convective heat transfer correlations inside tubes by analyzing both experimental and computational research. Convective heat transfer is influenced by fluid turbulence and boundary layers, with geometry significantly impacting flow conditions. Nusselt number correlations quantifying the heat transfer coefficient are vital for various applications. Previous reviews are summarized regarding nanofluid heat transfer modeling approaches in circular tubes. Additionally, comparisons of tube shapes like circular, elliptical and flat geometries are needed. Recent database searches indicate more experimental than numerical studies have been conducted, with greater focus on turbulent flows. This review systematically evaluates laminar and turbulent forced heat transfer correlations reported in the open literature, discussing both older established work as well as newer findings to provide a holistic perspective. The goal is to better understand convective transportation dependencies for optimized thermal design and performance improvement.

**Nomenclature**

| | |
|---|---|
| $D_h$ | Hydraulic diameter |
| $\varphi_v$ | Volume fraction |
| ANN | Artificial neural networks |
| b/D | Blockage ratio |
| CFD | Computational fluid dynamics |
| $D_n$ | Dean number |
| $f_{c,r}/f_{str,r}$ | Curvature ratio |
| FD | Fully developed |
| g/D | Gap spacing / tube diameter |
| Gr | Gershoff number |
| H | Hight |
| $k$ | Curvature ratio |
| Nu | Nusselt number |
| $P_r$ | Prandtl number |
| Re | Reynold number |
| $Re_{Dh}$ | Hydraulic Reynold number |
| $\delta$ | Thickness boundary layer |
| $Nu_{avg}$ | Average Nussle number |
| $f$ | Frictional factor |
| $\varepsilon$ | Relative roughness |
| $\mu$ | Viscosity |

## 1. Introduction

Convection heat transfer involves the transfer of heat between a solid surface and a fluid that comes into contact with it. Turbulence and the thickness of the boundary layer have a big impact on how much heat is exchanged. Fluids with turbulent, or mixing, flows transfer heat much more efficiently than laminar, or non-mixing, flows. Forced convection, where an outside source like a pump moves the fluid, is widely used in industry because it provides excellent heat transfer compared to natural convection from differences in temperature alone. Several fluid properties affect the rate of heat transfer, such as density, flow rate, conductivity, and heat capacity. However, the geometry that the fluid moves through may be the single biggest influence. The shape can drastically change turbulence and conditions near the wall, altering how much heat is removed or added. Understanding laminar and turbulent forced convection inside tubes is important for predicting performance in many applications that rely on heat exchange through a tube, such as car radiators. The Nusselt number is a common way to quantify the convection heat transfer coefficient, and it depends on whether the flow is laminar or turbulent and the thermal conditions on the surface.

Several studies have examined forced convection in circular tubes in crossflow configurations [1–4]. Ota et al. [5] experimentally studied elliptic tubes, finding local Nusselt number varies from circular tubes. Baughn et al.[6] experimentally evaluated single, tandem and bundled circular cylinders under uniform heat flux, observing Nusselt number (Nu) depends on row location. Stanescu et al. [7] experimentally optimized circular tube spacing in crossflow, following work by Bejan et al. [8] on free convection optimizations. For instance, Khan et al. [9] experimentally evaluated elliptical tubes with minor/major axis ratio of 0.33, finding increased heat transfer with airflow/waterflow. Studies by Sanitjai and Goldstein [10] experimentally characterized three flow zones around circular tubes in crossflow liquids/air. Chang and Mills [11] experimentally evaluated circular tube aspect ratio effects, finding mean heat transfer coefficient increases with lower aspect ratios. Tahseen et al.[12,13] and Ishak et al. [14] experimentally evaluated in-line and staggered flat tubes, observing Nu increases with Re and heat flux. Matos et al.[15] numerically and experimentally optimized elliptical/circular finned tubes, finding elliptical tubes increased heat transfer up to 19% versus circular. Ibrahim and Gomaa [16] numerically and experimentally studied elliptic bank tubes, finding best performance at small Re, attack angle, and axis ratios. Furthermore, Tahseen et al. [16] numerically evaluated staggered/in-line flat tubes between parallel plates, again observing increased Nusselt number with Reynolds number. Overall, these studies provided insights into crossflow heat transfer optimization accounting for tube geometry, configuration, flow properties and optimization parameters.

Cieśliński [17] comprehensively reviewed the studies of forced convection and nanofluids hydrodynamics characteristics inside various tubes such as smooth circular, and straight tubes which represent basic geometries in a shell-and-tube heat exchanger. Detail correlation equations have been proposed in order to calculate Nu and heat transfer coefficients for both laminar /turbulent flows, including nanofluid type, nanoparticles concentration, and Re range. In addition, several studies have simulated nanofluid heat transfer inside tubes using various modeling approaches. Behzadmehr et al. [18] studied the turbulent forced convection applying two-phase mixture model in a circular tube, representing the first use of this model for nanofluids. Their results showed higher accuracy than a single-phase model and agreed with the experimental correlation from Xuan and Li [19]. Moreover, Heris et al. [20] used a dispersion model to simulate laminar convection in a circular tube. Moreover, studies mentioned in refs [21,22] have investigated laminar convection using a Eulerian-Lagrangian approach. Additionally, Lotfi et al. [23] compared a single-phase model, two-phase mixture model and two-phase Eulerian model for nanofluids flow in a horizontal circular tube, finding the two-phase mixture model to be most precise against experimental data. Hussein et al. [24] reviewed computational simulations and experimental studies that have investigated using small solid particles (less than 100nm) as additives to liquids in order to improve the low heat transfer properties and hydrodynamic flow characteristics. The review indicated that correlation equations relating Nu, Re, concentration and nanoparticle diameter could be developed. Further comparisons of shapes like circular, elliptical and flat tubes are also needed to maximize heat transfer with minimal pressure penalty.

Figure 1 shows the reported studies of laminar and turbulent correlations of forced heat transfer convection according to Scopus database. The figure classifies whether the study is laminar or turbulent or experimental or numerical approaches. It can be noticed that there are more experimental studies than the numerical.

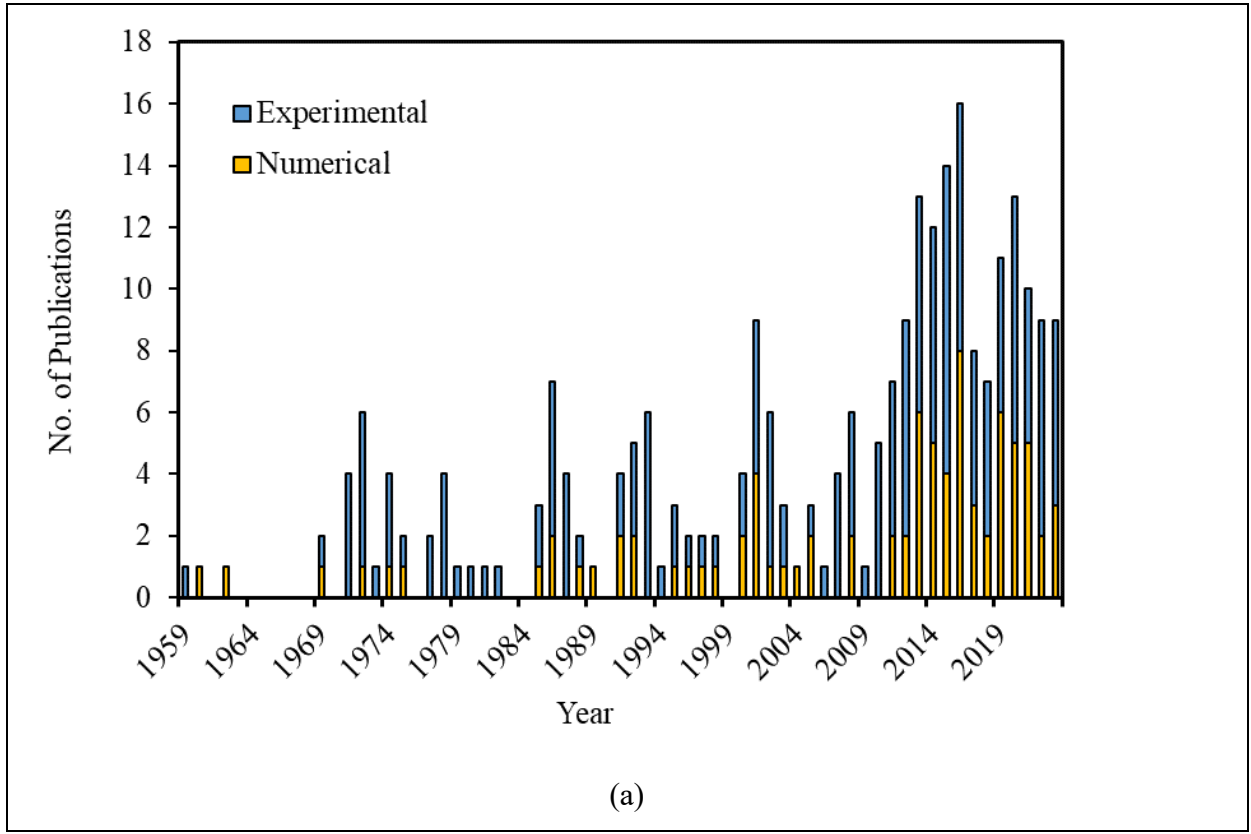

(a)

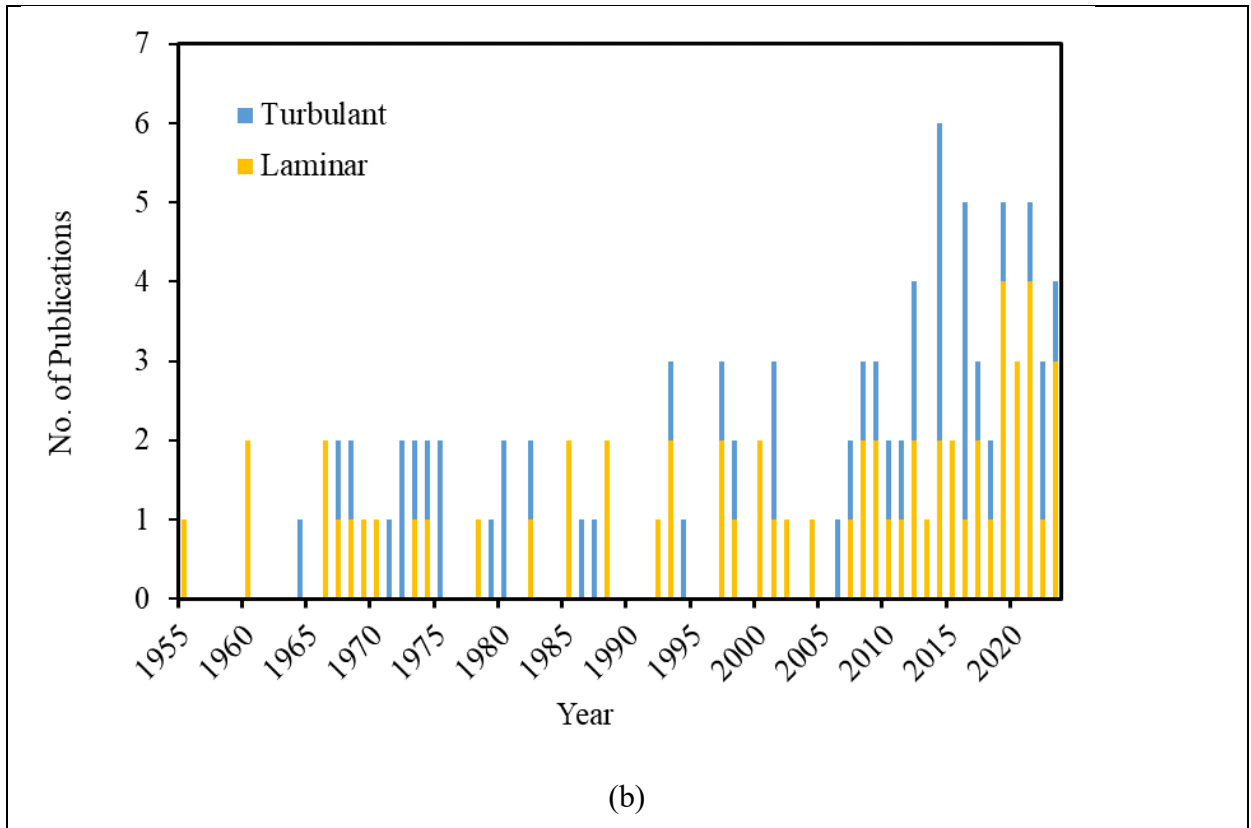

Figure 1. Reported studies of forced heat transfer convection correlations in the open literature from Scopus data; (a) Numerical and experimental studies (b) laminar and turbulent studies

In this work, a comprehensive review of laminar and turbulent correlations of forced heat transfer convection inside tubes is investigated. The review structure is divided into laminar and turbulent sections. Experimental and numerical studies of laminar and turbulent forced convection flows will be discussed and investigated. Although most of this topic studies and correlations were investigated a long time ago, this review will mention the most recent studies as possible. The uniqueness of this work is that it reviews the topic comprehensively for different types of topics and applications.

## 2. Laminar forced heat transfer convection inside tubes

For laminar flow in tubes, convection can be forced or mixed. Mixed convection involves buoyancy effects from radial density differences in the fluid. Distinguishing among forced and mixed convection is essential due to their associated Nu differ significantly [25]. Everts and Meyer [26] have mapped the horizontal tubes flow regime under constant heat flux to determine if conditions are forced or mixed. Metais and Eckert [27] similarly mapped vertical flow for constant heat flux and surface temperature boundaries. Most heat transfer textbooks [28–31] report the fully developed (FD) laminar Nusselt number under constant heat flux is a constant 4.36, independent of Re or Prandtl numbers (Pr). This value assumes constant fluid properties. In case of mixed convection situations, Nu can be 180-520% higher than 4.36 due to additional buoyancy effects on heat transfer not considered in the theoretical 4.36 value derived for purely forced convection [32–34]. Properly characterizing the flow regime is thus critical to predicting heat transfer performance.

This section reviews and discusses the experimental and numerical studies that studied the laminar forced convection correlations in detail.

## 2.1. Experimental studies

Experimental studies of forced convective flow inside smooth horizontal tubes are challenging due to the need to minimize buoyancy effects on heat transfer. One such study investigated this issue using water, which has a Prandtl number ranging from 3.5-8.1 [35]. The test setup varied the Reynolds number from 400 to 6000 under constant heat flux ranged from 1 to 8 kW/m². The results showed that at Reynolds numbers above 1000, for FD laminar forced convection, Nu is not a constant value of 4.36 as previously derived, rather, it will vary as a function of the Re when Re exceeded 1000. In other words, at higher Reynolds numbers of buoyancy forces can no longer be ignored and influence heat transfer, so the classic equation giving a constant Nusselt number is invalid in those flow conditions inside horizontal tubes. This highlights the challenges of experimental studies in this regime due to buoyancy effects on heat transfer coefficients. The derived correlation is:

$$Nu = 4.36 + 5.36 * 10^{-9} Re^{2.39}$$

The above correlation can be used for Re of 600 and 3000 if transition does not occur (2100–2300) [35]. As can been seen from Figure 2, the units of the experiment are fitted with the curve along the variation of the Re.

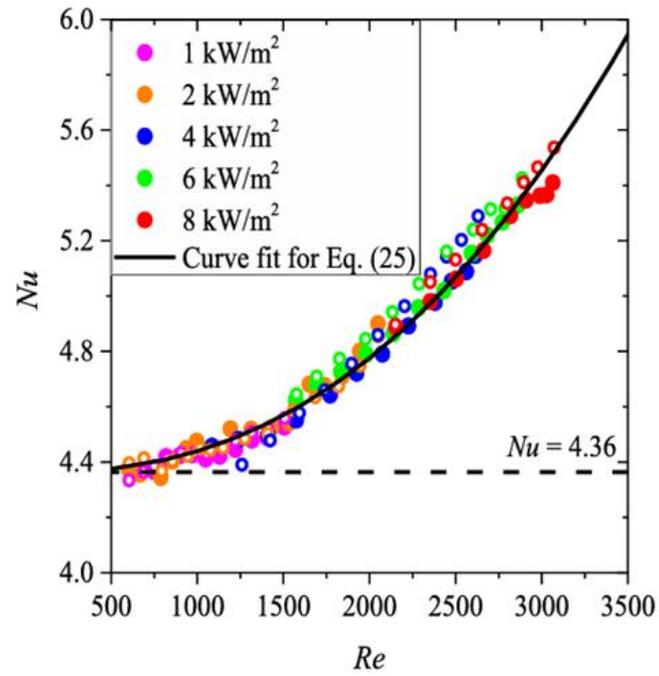

Figure 2. A linear regression analysis of the average laminar forced convection Nu obtained from the experiments to the corresponding Re [35].

Ghajar et al. [36] conducted an experimental test in a circular straight tube with constant heat flux conditions. The study separated the modes of heat transfer in the laminar regime into forced and mixed convection. Figure 3 shows the proposed setup.

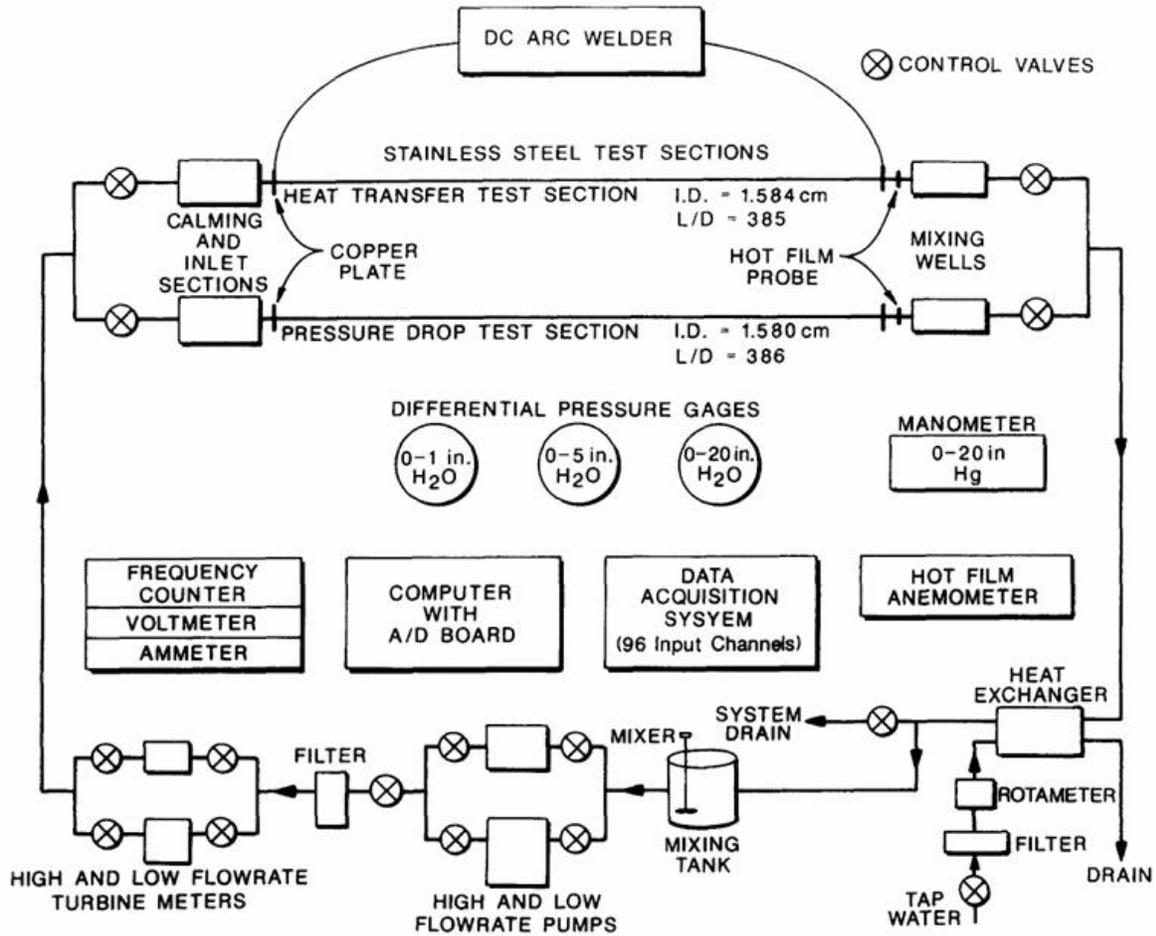

Figure 3. Experimental setup diagram proposed in ref [10].

As noticed, the study observed that near the entrance of horizontal tubes, the heat transfer behavior was ruled by pure forced convection. The reason was because secondary flows, influenced by buoyancy, required a certain development length before impacting heat transfer. To take into consideration the regions of entry and completely established laminar flow, the researchers proposed a correlating equation as [36]:

$$Nu_l = 1.24 \left[ \left( Re\, Pr \frac{D}{x} \right) + 0.025 (G_r P_r)^{0.75} \right]^{\frac{1}{3}} (\mu/\mu_w)^{0.14}$$

This is valid when:

$$3 \leq \frac{D}{x} \leq 192, \quad 280 \leq Re \leq 3800, 40 \leq P_r \leq 160,$$

$$1000 \leq G_r \leq 2.8 \times 10^4, \quad 1.2 \leq \frac{\mu}{\mu_w} \leq 3.8$$

The above equation is valid under the above conditions. In addition, it can be utilized to all three inlets and is valid for both entry as well as fully developed zones' laminar driven and mixed convection [36]. Most of the experimental studies concluded by using either theoretical models or empirical based. However, this study is aimed to compare different correlating methods using Artificial Neural Networks (ANN). This comparison will be investigated by using heat transfer data of laminar and turbulent region that were obtained from an old single-phase heat transfer study [36]. The below correlation predicts more data precisely compared to the traditional correlation [37]

$$Nu = 10.49 - 0.149 * \left(\frac{\mu_b}{\mu_w}\right)^2 - \left(Gz * \left(\frac{\mu_b}{\mu_w}\right)\right)^{0.5}$$

A study examined heat transfer for laminar fluid flow over flat tubes placed within a channel [38]. The working fluid used was air, as at five velocities ranging from 0.2 to 1.0 m/s, corresponding to Reynolds numbers based on hydraulic diameter between 124.5 to 622.5. The flat tube surface received a uniform heat flux. Results from the experiment produced a correlation relating average Nusselt number to pertinent parameters. This correlation achieved good agreement with measured data, accurately predicting values with a minimum root mean square error of 99.70%. the mathematical expression is stated as:

$$Nu_{Dh\,Ave} = C * Re_{Dh}^m$$

Where c and m are constant depend on the heat flux.

Another experimental study is conducted to investigate the overall heat transfer coefficients of 3 small spheres in different forced flow types, air, water, and oil [39]. As a result, the Nu number is stated as below:

$$Nu = 2.0 + 1.3\,(Pr)0.15 + 0.66\,(Pr)0.31\,(Re)0.50$$

This correlation is valid for:

$$0.4 \leq Re \leq 2100$$

$$0.71 \leq P_r \leq 380$$

The study validated that the heat transfer behavior for the cylindrical heating element orientation could be accurately predicted using the given correlation developed from experimental measurements in that flow arrangement as shown in the expression below [39]:

$$Nu = 0.42\,(Pr)^{0.20} + 0.57\,(Pr)^{0.33}\,(Re)^{0.50}$$

Tahseen et al. [13] experimentally investigated laminar forced convection flows inside flat tubes. Tests were conducted at five air velocities from 0.2-1.0 meter/s, corresponding to Re based on (ReDh) of 124.5-622.5. Uniform heat fluxes of 354.9, 1016.3 and 1935.8 W/m2 were supplied at the tube surface (see Figure 4). Results showed average Nus for the flat tube increased with hydraulic Re at fixed heat flux. Average Nu also rose with increasing heat flux at fixed ReDh. Additionally, they found that, as the free stream velocities increased, the friction factor decreased. power law relationship was found between average Nu and ReDh. The derived Nuavg-ReDh correlation accurately predicted measured data with a minimum RMSE of 99.70%.

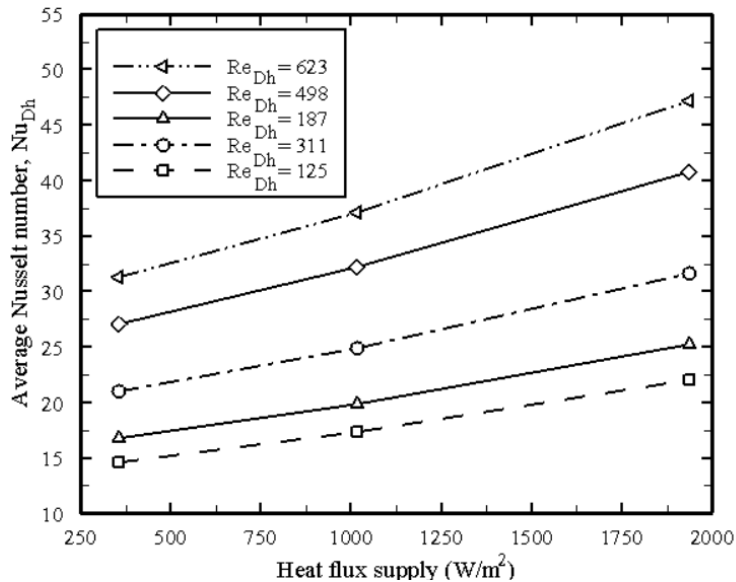

Figure 4. The relation between heat flux and average Nu for different hydraulic Re conducted in study Ref [13].

Conduction experimental forced convection tests in smooth circular horizontal tubes are challenging due to the need to minimize buoyancy effects on heat transfer through use of very low heat fluxes. Hence, a study experimentally investigated pressure drop and forced convection performance using vertical upward and downward water flow (Prandtl 3.5-8.1) in a smooth circular test section [25]. The experimental set up is presented in Figure 5. Experiments varied Reynolds numbers from 400-6000 and heat fluxes from 1-8 kW/m2. While flow directions and heat fluxes had no influence in laminar flows, the FD laminar forced convection Nu was found to vary with Re above 1000 rather than remaining constant at 4.36. A revised laminar Nu correlation was developed. Friction factors matched the expected fully developed laminar value of 64/Re.

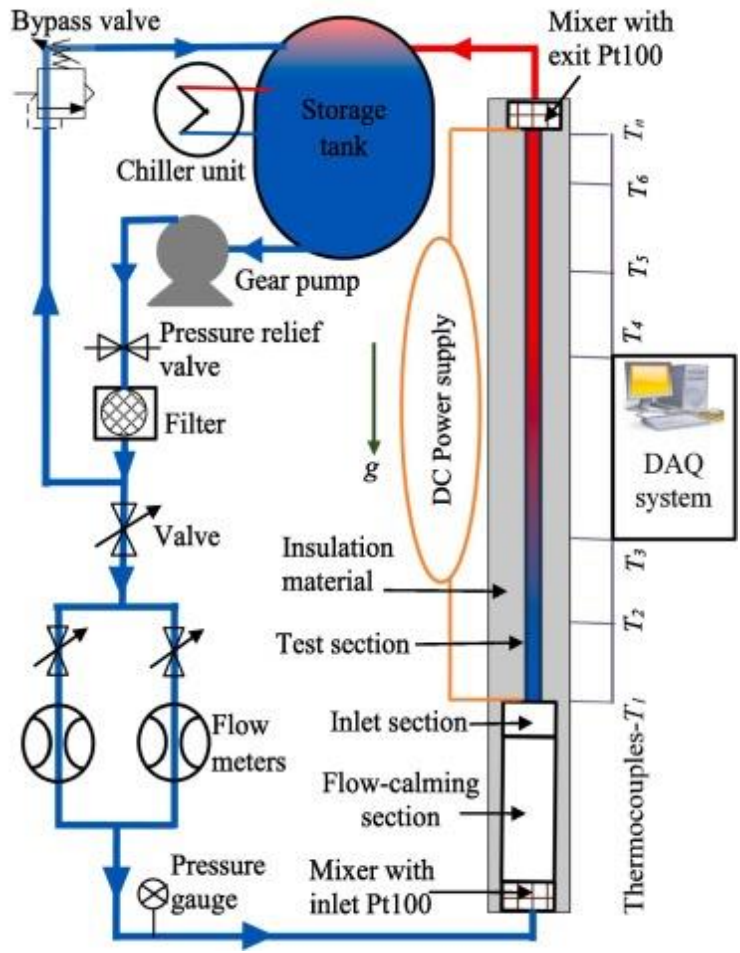

Figure 5. Experimental set up proposed in ref study [25].

## 2.2. Numerical studies

Nanoparticles have significant potential to enhance heat transfer in various applications. As a result, researchers are actively studying the effects of nanoparticles and developing tailored correlations. This study experimentally investigated forced convection heat transfer using nanofluids inside uniformly heated tubes and between parallel discs. The nanofluids tested were water-Al2O3 and ethylene glycol-Al2O3 mixtures [40], with nanoparticle concentration varied. Results clearly showed the affect inclusion and concentration of nanoparticles on heat transfer. Ethylene glycol-Al2O3 nanofluid provided greater improvement compared to the water-based nanofluids. Correlations were derived from the experiments to compute Nusselt number as a function of Reynolds and Prandtl numbers for the nanofluid systems.

For constant heat flux:

$$Nu_{avr} = 0.086 Re^{0.55} P_r^{0.5}$$

For constant surface temperatures:

$$Nu_{avr} = 0.28 Re^{0.35} P_r^{0.36}$$

Another study used CFD to analyze the convective influence of nanofluid in the tube developing region under constat heat flux. The fluid is a mixture of water and Al2O3 [41]. The average sizes of particles are 45 and 150 nm. In addition, the concentrations of the four particles that have been used are 1, 2, 4 and 6 wt.%. to analyze the size influence, the study conducted at different Re that various between (500 < Re < 2500) at different axial location of the tube. Based on simulation outcomes, the Nusselt number equation is derived as a function of dimensionless numbers. The results of the derived equation are complied with experimental result with a maximum error around 10%. The equation is stated as following:

$$Nu = 2.03 p_v^{0.06} \left(\frac{x}{D}\right)^{-0.37} Re^{0.293} P_r^{0.6}$$

Additionally, a numerical study investigated the nanofluid influence on enhancing the heat transfer using mixture of $FE_3O_4$ and water as work fluid [42]. This study fixed the dimensions of the channel and length. However, different geometries of cross section are investigated under different

Re numbers that varied from 500 up to 2000. In addition, the percentage of nanoparticles concentration is between 1% and 5%. The findings showed that the nanofluid particle enhanced the heat transfer inside the channel whatever the geometry is. Cylindrical geometry gave the highest heat transfer value while the tringle geometry gave the lowest. However, the highest ratio of heat transfer increment was given by the tringle geometry. a new Nu correlation equation is developed as function of the Re number, the volumetric nanoparticle concentration and some constants has been proposed for different geometries:

$$Nu = a * Re^b + c * \varphi$$

Where $a$, $b$ and $c$ are constants and they can be found in Table 1:

Table 1. constant values of $a$, $b$ and $c$ for different tube cross sections mentioned in ref [42].

| Tube type / constant | a | b | c |
|---|---|---|---|
| Tringle | 0.282754 | 0.40213 | 3.55865 |
| Square | 0.539653 | 0.347802 | 4.23371 |
| Cylindrical | 0.873747 | 0.312881 | 4.98564 |
| Rectangular | 0.556318 | 0.347361 | 4.36119 |

Additionally, Davarnejad et al. conducted a CFD analysis to understand the $Al_2O_3$ nanoparticles effect on laminar convective heat transfer [43]. The results from the simulation indicated that heat transfer was highest with greater nanoparticle concentrations. Moreover, the relation between Peclet number and heat transfer coefficient was estimated to be directly proportional, with increases in Peclet number correlating to higher heat transfer coefficient. The simulated data was then validated by comparing it to experimental data from a separate study, showing good agreement with less than 10% discrepancy. From the simulated results, a new heat transfer correlation was derived relating Nusselt number as a function of several important parameters including the Re that ranges between 700 and 2050, Pr, particle volume fraction, and axial locations within the pipe defined as the horizontal distance divided by pipe diameter. The equation is stated as following:

$$Nu = 0.18665 \emptyset^{-0.00728} \left(\frac{x}{D}\right)^{0.1036} Re^{0.368718} P_r^{0.3992}$$

Moreover, Kendoush et al. conducted a study to analyze the influence of low Prandtl number fluids on heat transfer from isothermal spherical particles [44]. Convective heat and mass transmission over a laminar, incompressible boundary layer a stationary isothermal surface have been predicted analytically. It was assumed in the investigation that there was an inviscid flow outside the viscous boundary layer. The investigation of the heat and mass transfer rates at the forward and rear stagnation points, as well as the local and total transfer rates in the front and wake zones, led to the development of new correlations, for low Prandtl number fluids interacting with the spherical particle surface under laminar flow conditions. Comparing the results with experimental data, the correlation showed good results. The correlation is stated below:

$$Nu = 1.13(RePr)^{0.5}$$

A numerical 3D laminar flow and heat transfer investigation used two types of nanofluids, $Al_2O_3$ and CuO mixed with water, circulating through a flat tube representative of an automotive radiator in ref [45]. The aim was to evaluate the nanofluids' performance relative to the base fluid. Findings indicated the nanofluids provided good enhancement of heat transfer through the medium compared to water alone. Additionally, the study developed a Nusselt number correlation based on its numerical simulations to predict heat transfer for these nanofluid systems flow through the representative radiator tube geometry.

$$Nu_{avg} = 1.9421 \left( Re \, P_r \frac{D_h}{Z} \right)^{0.3} \quad for \; Re \, P_r \frac{D_h}{Z} \, 33.33$$

$$Nu_{avg} = 6.1 + 0.003675 \left( Re \, P_r \frac{D_h}{Z} \right) \quad for \; Re \, P_r \frac{D_h}{Z} < 33.33$$

Numerical studies of laminar forced convection producing flow in different straight canal shapes, such as a round tube with fixed surface temperature, were carried out by Nonino et al. [46]. When other parameters were kept constant, it was discovered that taking into consideration how viscosity changes with temperature along the channel's length had an important effect on laminar forced convection heat transfer coefficients. Similarly, Nouar [47] discovered that decreased viscosity caused by increased temperature throughout the tube length affected heat transfer coefficients. Zhai et al. [48] studied how the effects of varying fluid properties, such as axial conduction, impacted laminar forced convection heat transfer coefficients in tubes. The researchers found that changes in fluid characteristics had a greater influence on the velocity profile in entrance regions of the tubes. Meanwhile, temperature profiles within the tubes were affected more significantly in sections where the fluid flow was fully developed. So in entrance sections, velocity was more sensitive to fluid property changes, while in fully developed regions, temperature was more influenced by such changes, according to their findings. Overall, these studies demonstrated the importance of considering variable fluid properties like temperature-dependent viscosity for accurately predicting laminar forced convection heat transfer performance, particularly in developing flow and entrance regions.

## 2.3. Summary of Laminar correlations

To sum up, the below table summarized all the correlations that collected from different studies for laminar forced convective heat transfer.

| Important notes | Correlations | Study approach | Ref. |
|---|---|---|---|
| Single-phase study | $Nu = 4.36 + 5.36 * 10^{-9} Re^{2.39}$ | Exp. | [49] |
| The tube is circular with three different inlet configurations | $Nu_l = 124 \left[ \left( Re\, P_r \frac{D}{x} \right) + 0.025 (G_r P_r)^{0.75} \right]^{\frac{1}{3}} (\mu/\mu_w)^{0.14}$<br>This is valid for:<br>$3 \le \frac{D}{x} \le 192, \quad 280 \le Re \le 3800, 40 \le P_r \le 160,$<br>$1000 \le G_r \le 2.8 \times 10^4, \quad 1.2 \le \frac{\mu}{\mu_w} \le 3.8$ | Exp. | [50] |
| The tube is circular with three different inlet configurations utilizing (ANN) | $Nu = 10.49 - 0.149 * \left(\frac{\mu_b}{\mu_w}\right)^2 - \left(Gz * \left(\frac{\mu_b}{\mu_w}\right)\right)^{0.5}$ | Exp. | [37] |
| Flat tube was used. | $Nu_{Dh\,Ave} = C * Re_{Dh}^m$ | Exp. | [38] |

| spheres to flowing media | $Nu = 2.0 + 1.3 \times (Pr) \times 0.15$ $+ 0.66 \times (Pr) \times 0.31\,(Re) \times 0.50$ This correlation is valid for: $0.4 \leq Re \leq 2100$ $0.71 \leq P_r \leq 380$ | Exp. | [39] |
|---|---|---|---|
| The tube was uniformly heated and used nanofluid particles | For constant heat flux: $Nu_{avr} = 0.086 Re^{0.55} Pr^{0.5}$ For constant wall temperature: $Nu_{avr} = 0.28 Re^{0.35} Pr^{0.36}$ | Numerical | [40] |
| The study investigated the developing region inside the tube | $Nu = 2.03 p_v^{0.06} \left(\dfrac{x}{D}\right)^{-0.37} Re^{0.293} P_r^{0.6}$ | CFD | [41] |
| A hemispherical porous tube with a uniform magnetic field | $Nu = a * Re^b + C * \varphi$ valid for $500 \leq Re \leq 2000$ and $0 \leq \Phi \leq 0.05$ | Numerical | [42] |
| Nanofluid in a Circular Cross-section | $Nu = 0.18665 \emptyset^{-0.00728} \left(\dfrac{x}{D}\right)^{0.1036} Re^{0.368718} P_r^{0.3992}$ | Simulation | [43] |
| The study used isothermal spherical particles | $Nu = 1.13 (RePr)^{0.5}$ | Analytical | [44] |
| Th study investigated flat tube of a radiator | $Nu_{avg} = 1.9421 \left(Re\, P_r \dfrac{D_h}{Z}\right)^{0.3}\ for\ Re\, P_r \dfrac{D_h}{Z}\ 33.33$ $Nu_{avg} = 6.1 + 0.003675 \left(Re\, P_r \dfrac{D_h}{Z}\right)\ for\ Re\, P_r \dfrac{D_h}{Z} < 33.33$ | Numerical | [45] |

## 3. Turbulent forced heat transfer convection correlation inside tubes

This section reviews and discusses the experimental and numerical studies that studied the turbulent forced convection correlations in detail.

### 3.1. Experimental studies

Helical tubes are often used in industrial applications due to benefits like accommodating thermal expansion, a compact design, and improved heat transfer performance. However, fluid flow within helical tubes is more complicated than in straight tubes. This is because secondary flows are generated in the cross-section due to the coiling, as studies have shown. The maximum velocity shifts towards the outer wall farthest from the helical axis, resulting in asymmetric velocity and temperature profiles across the section [51,52]. Additionally, shear stress and heat transfer coefficients around the tube's circumferential wall become uneven. The flow field itself is stabilized by centrifugal forces, which causes the transition from laminar to turbulent flow to be delayed compared to straight tubes [53]. These factors associated with the curved, rotating geometries of helical tubes complicate the internal fluid dynamics versus simple straight pipes [53]. While centrifugal forces still influence turbulent flow in helical tubes, their dominance decreases compared to laminar flow [54]. Several empirical friction factor correlations have been derived from experimental regression analyses, with limited theoretical equations. For instance, Ito [54] proposed two correlations using power law and logarithmic velocity distributions that are widely and accurately used [55], though underestimate friction at high Reynolds numbers over 105. Zhou and Shah [56] attributed such discrepancies to neglected roughness effects at commercially high flows. A new correlation accounting for roughness through additional regression of tests in rough helical tubes have been developed by Zhao et al. [57], validated up to curvature ratios of 0.063. They found roughness impacts helical flows more than straight pipes. Despite review papers on helical convection correlations, roughness influences remain underdiscussed despite established commercial tube significance at higher Reynolds numbers [53]. Figure 6 represents the correlations among the critical Re and the helical tube curvature ratio calculated by several studies.

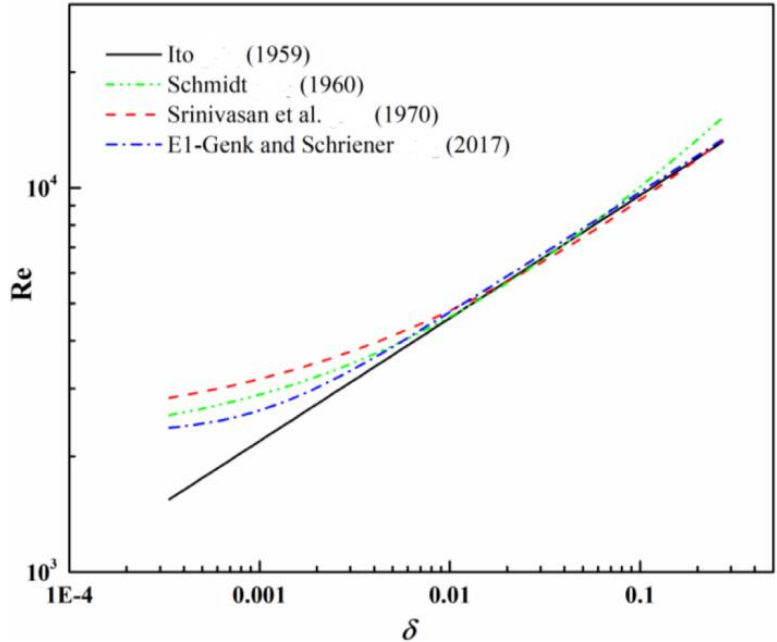

Figure 6. Critical Reynolds number correlations vs curvature ratio calculated by different studies [54,58–60]

Aicher et al. [49] examined mixed natural and forced convection involving both upward and downward flow in vertical tubes experimentally. The study proposed new insights that examined the effects of varying length-to-diameter ratios and opposing or concurrent heat and mass flux directions on heat transfer within vertically oriented tubes. Two new empirical correlations were developed that improved upon existing models by accounting for the laminarization phenomenon where turbulence is diminished during aiding mixed convection. Specifically, the correlations more accurately represented the full dataset, predicting measured values from both the current study and literature with over 80% accuracy by incorporating parameters not considered in past equations. This captured the influence of factors like length-to-diameter ratio and directionality of heat/mass fluxes that prior relationships did not address. A lot of research investigations in the open literature have been tested experimenting the turbulent forced heat transfer convection flow of nanofluid material inside tubes with a Reynold number ranging from 2300 [61–66]. Many of them used Dittus–Boelter equation of pure water which is expressed as [24]:

$$Nu = 0.023\, Re^{0.8} P_r^{0.4}$$

Azmi et al. [67] experimentally measured heat transfer coefficients for SiO2/water and TiO2/water nanofluids with volumes up to 3% flowing turbulently in a circular tube. Tests were run at Reynolds numbers from 5000-25000 and a bulk temperature of 30°C. Twisted tape inserts with twist ratios ranging from 5-93 were used inside the tube. Heat transfer enhancement was found to decrease with greater twist ratios. At 3.0% volume, a ratio of five produced a 27.9% higher coefficient for SiO2/water nanofluid versus water alone, but only 11.4% higher for TiO2/water nanofluid. Correlations were developed to estimate the Nusselt number for both water and nanofluid turbulent flow with twisted tapes, based on the test conditions. This provided predictive equations for assessing heat transfer performance (see Figure 7).

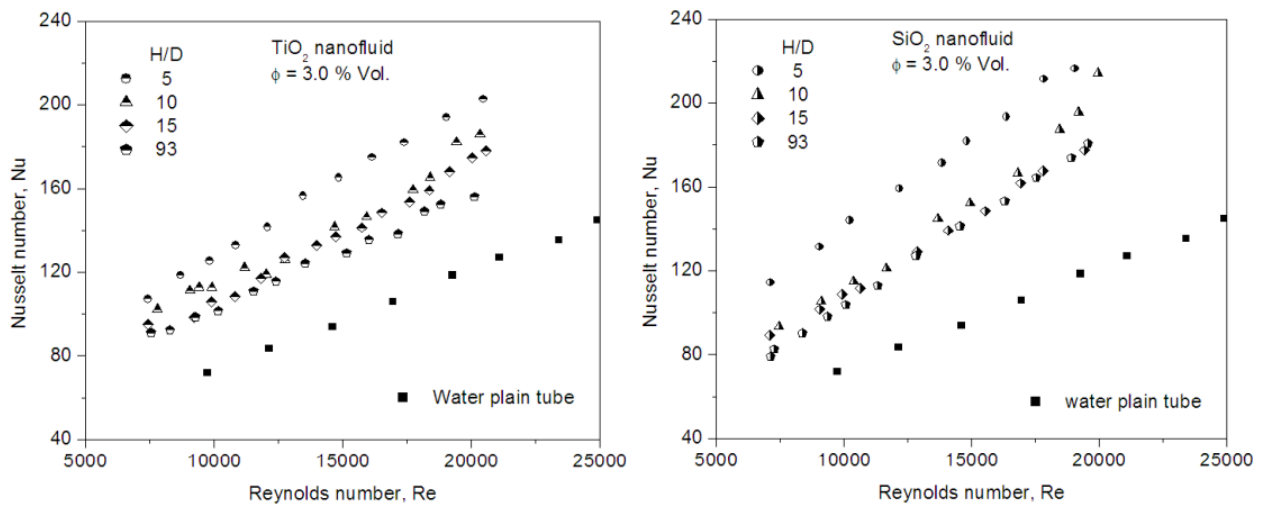

Figure 7. Nu vs Re of nanofluid at 3% volume concentration of TiO2 and SiO2 nanofluids [67].

Pak et al. [68] conducted an experiment that investigated turbulent friction and heat transfer using water suspended with 13nm γ-alumina and 27nm titanium dioxide nanoparticles between 1-3% volume concentrations. Viscosity measurements significantly diverged from classical suspension predictions. Friction factors agreed well with Kay's single-phase correlation. The Nusselt number rose with concentration and Reynolds number, though heat transfer coefficients for 3% concentration fluids were slightly less than water at constant velocity. Based on these findings, the study recommended selecting larger, higher thermal conductivity particles to enhance heat transfer performance of dispersed fluids. The study proposed a new correlation to predict turbulent convective heat transfer for diluted dispersed fluids containing submicron metal oxide nanoparticles. Additionally, several other experimental studies investigated nanofluid heat transfer and pressure drop characteristics. Some studies found minimal enhancement, while others observed higher heat transfer coefficients that increased with nanoparticle concentration but were unaffected by changes in Reynolds number. Sonawane et al. [69] found maximum heat transfer enhancement of 47% for 50nm Al2O3 in aviation fuel, versus 36% for 150nm particles. Abbasian and Amani [70] examined TiO2-water and different nanoparticle sizes, finding 20nm produced the highest Nusselt numbers and pressure drops as indicated in Figure 8. In summary, these experimental investigations provided insight into how nanofluid thermophysical properties, heat transfer performance and pressure characteristics are influenced by factors like base fluid, nanoparticle material, size, concentration, and flow conditions.

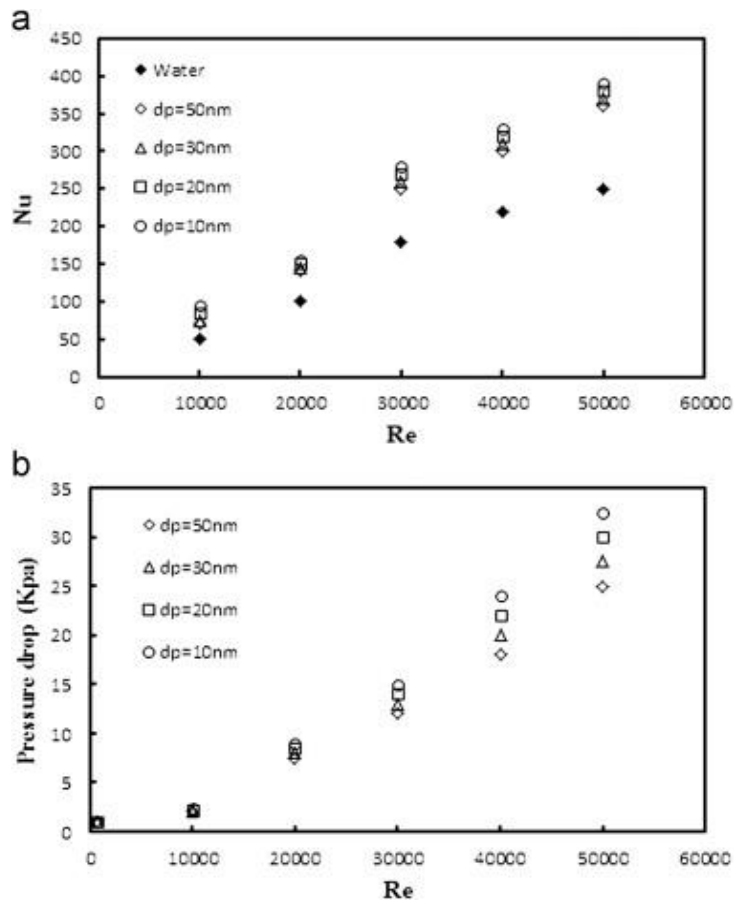

Figure 8. The effect of diameter size on the Nussle number and pressure drops [70].

In another study [71], forced convection from the outer surfaces of helical coiled tubes under a constant wall heat flux condition was experimentally analyzed. Ten tube samples with varying diameter ratios from 7.086 to 16.142 and pitch ratios from 1.81 to 3.205 were tested in an open-circuit wind tunnel operating in suction mode over a Reynolds number range of $6.6 \times 10^2$ to $2.3 \times 10^3$. The results showed diameter and pitch ratios significantly impacted the average heat transfer coefficient. At constant Reynolds and pitch ratios, average Nusselt number increased with diameter ratio, and at constant Reynolds and diameter ratio, Nusselt number rose with pitch ratio. Based on the experimental findings, an empirical correlation was derived relating Nusselt number to Reynolds number, diameter ratio, and pitch ratio.

Figure 9 shows the effect of pitch ratio (P/do) on average Nusselt number, demonstrating Num increases with both increasing Reynolds number and P/do. Specifically, at a constant Re of 1642, Num was 37% higher for a P/do of 3.205 compared to a P/do of 1.181. Meanwhile, at constant P/do of 2.181, Num increased 31.4% when raising Re from 660.3 to 2235. The physical mechanism is that higher pitch ratios enhance heat transfer by raising the coil height and distances between turns, allowing greater airflow both between coil loops and within the coil core. This leads to augmented convection as the increased airflow amplifies heat transfer coefficient values along the helical surface. Therefore, pitch ratio emerges as another important design factor impacting heat transfer performance.

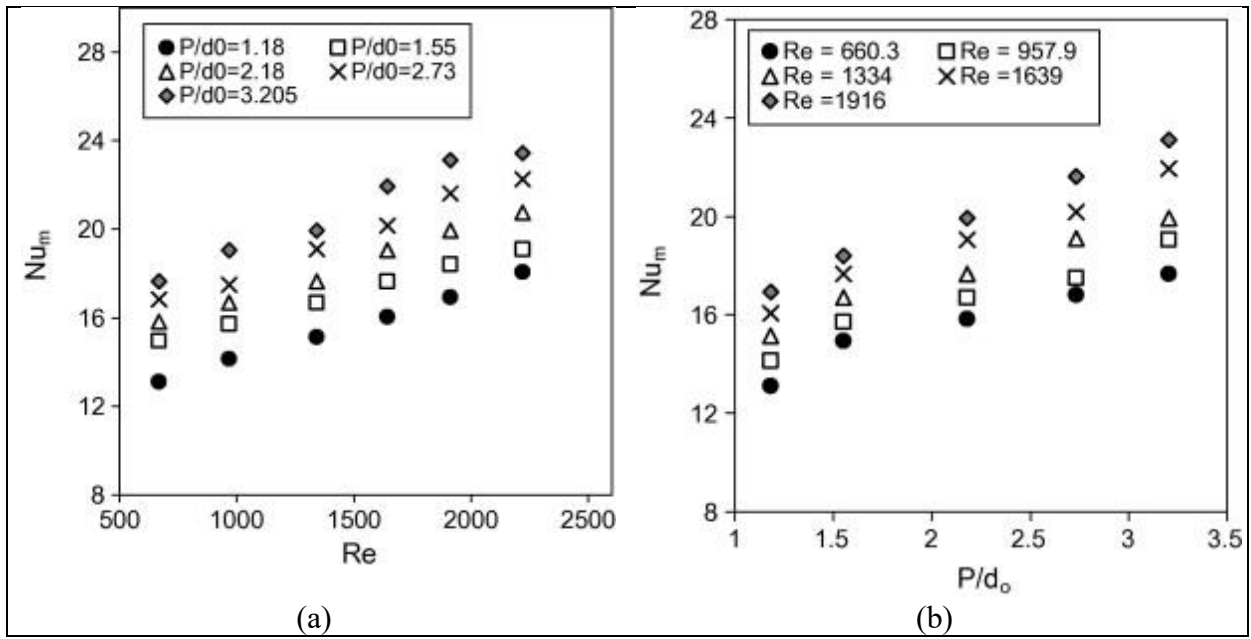

Figure 9. Nu varied with a) Re at different pitch to diameter ratios and a diameter to hydraulic diameter ratio of 16.142 and heat flux of 60 W/m2. b) Pitch to diameter ratio at different Re and a diameter to hydraulic diameter ratio of 16.142 and heat flux of 60 W/m2. [71].

## 3.2. Numerical studies

Several reported studies have numerically simulated the turbulent forced convection inside different type of tubes for different applications [72–75]. For instance, recently, Jedsadaratanachai et al. [74] numerically analyzed the performance of heat transfer, pressure drop, and thermal enhancement in a circular tube heat exchanger fitted with V-shaped orifices. The analyses examined the effects of blockage ratio, gap spacing ratio, and orifice arrangement for turbulent flow with Reynolds numbers of 3000-10,000. The results show that the gap spacing ratio was the primary determinant of changes to flow and heat transfer patterns. Adjusting the gap distance helped optimize thermal performance, especially at higher blockage ratios. Additionally, the maximum thermal performance of the system occurred at a blockage ratio of around 2.25 for the lowest Reynolds number of 3000 (see Figure 10).

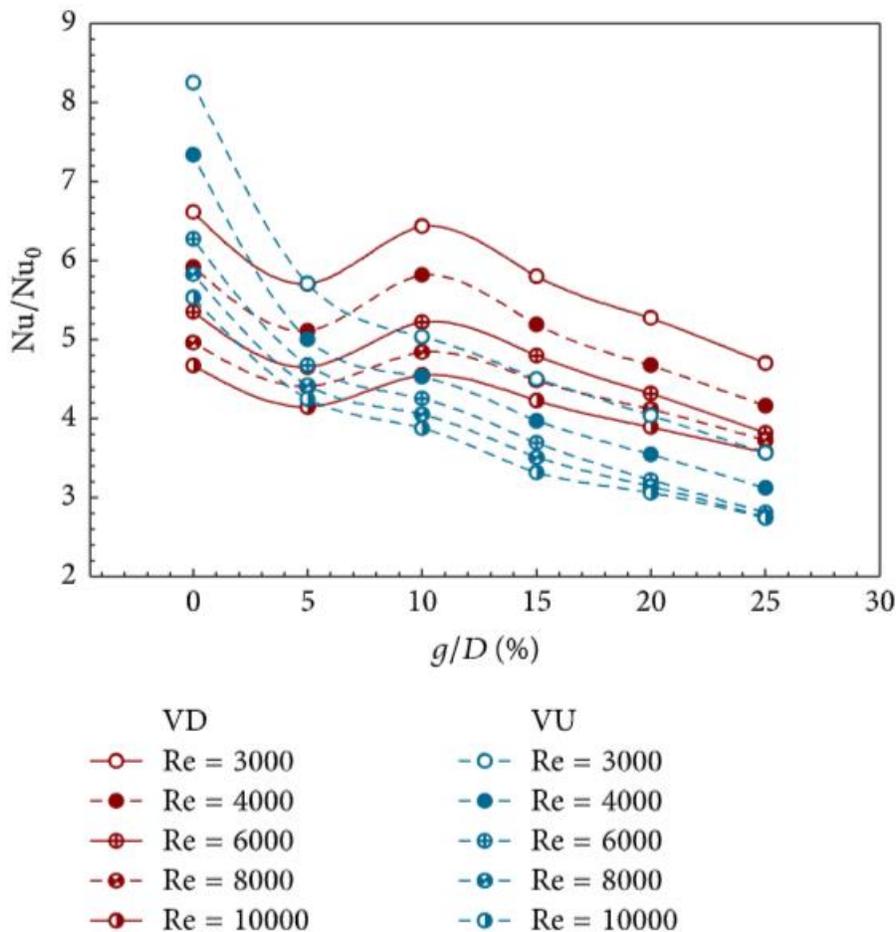

Figure 10. the effect of Nu /Nu0 on gap spacing / tube diameter (g/D) for blockage ratio of 0.2 adopted from Ref [74].

The study by Lai et al. [76] identified three pairs of vortices in turbulent curved tube flow patterns. One pair was the typical Dean vortices. Another pair formed in the tube core due to pressure differences. A third pair appeared near the outer wall driven by turbulence. Additionally, Ivan and Michele [77] computationally modeled turbulent curved tube flows using various turbulence models. Their results found excellent agreement between pressure drop predictions and experimental data when using the SST K-ω and RMS K-ω models. However, simulations with the standard k-omega model employing wall functions produced unsatisfactory results compared to measured pressure drops.

Flow and heat transfer in helical tubes is more complex than straight tubes due to secondary flows from centrifugal forces, depending on parameters like Reynolds number, curvature ratio and pitch. Zhao et al. [53] experimentally measured friction factors in helical tubes to validate correlations. They recommended friction factor correlations covering wide Reynolds number and curvature ratio ranges for both laminar and turbulent regimes. For commercial heat exchangers with Reynolds over 105, the study suggested considering roughness effects per another reference. Additionally, new Nusselt number correlations for laminar and turbulent convection in helical tubes were developed by regressing experimental data relating parameters that influence heat transfer performance in helical geometries [53]. Figure 11 represent the effect of curvature ratio on the friction factor for different Reynold number [57]. Findings presented in Figure 11 indicated that as the roughness height increases, so too does the curvature ratio ($f_{c,r}/f_{str,r}$). This indicates that the impact of curvature ratio on friction factors is augmented by greater surface roughness inside rough helical tubes. Specifically, when the tube walls become rougher as measured by increasing roughness height, the relative effects of curvature, as represented by the variable $f_{c,r}/f_{str,r}$, are also increased on frictional loses. In other words, rougher inner pipe surfaces in helical tubes serve to enhance the influence of curvature ratio changes brought on by the helical coil shape. Therefore, the findings demonstrate that accounting for curvature ratio effects needs to be considered pipe roughness, as rougher walls strengthen the sensitivity of friction factors to variations in tube geometrical curvature [57].

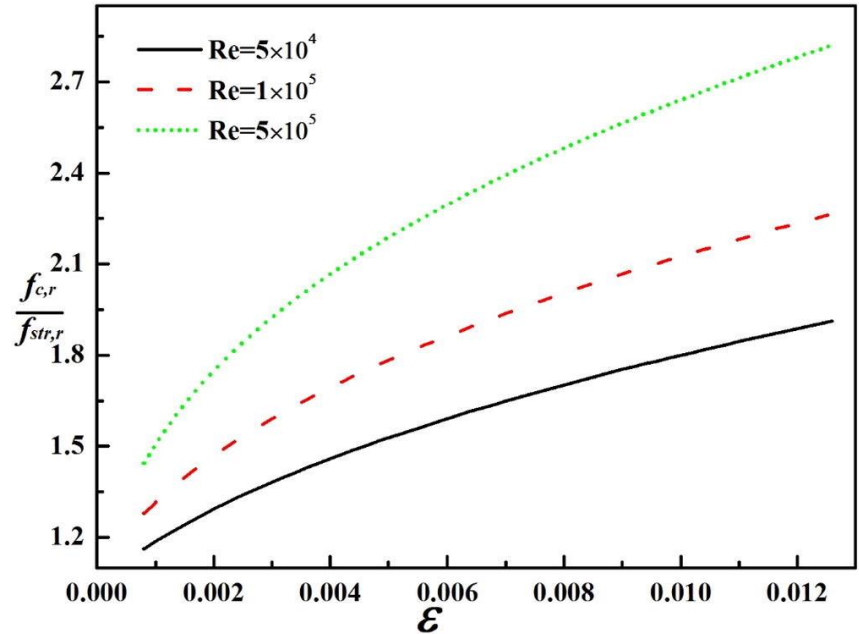

Figure 11. The relation between curvature ratio and friction ratio with different Re numbers [57].

Mahmoudi et al. [78] experimentally and numerically analyzed forced convection and pressure drop using TiO2/water nanofluid inside helically coiled tubes (see Figure 12). They tested Reynolds numbers from 3000-18,000 and nanoparticle concentrations of 0.1-0.5% for five curvature ratios. The numerical model considered thermophysical property variation with nanofluid temperature and concentration, using the k-ε model for turbulent flow simulation. Results showed up to a 30% increase in Nu when using nanofluid versus water. Correlations were introduced to determine the average Nusselt number and friction factor in helically coiled tubes under constant wall temperature for both laminar and turbulent flows. This provided insights into physical properties effect, flow characteristics, and geometric factors on heat transfer enhancement and pressure penalties associated with nanofluid flow in helical pipes. The aim was to provide additional insight into geometry and property impacts on augmentation and drops.

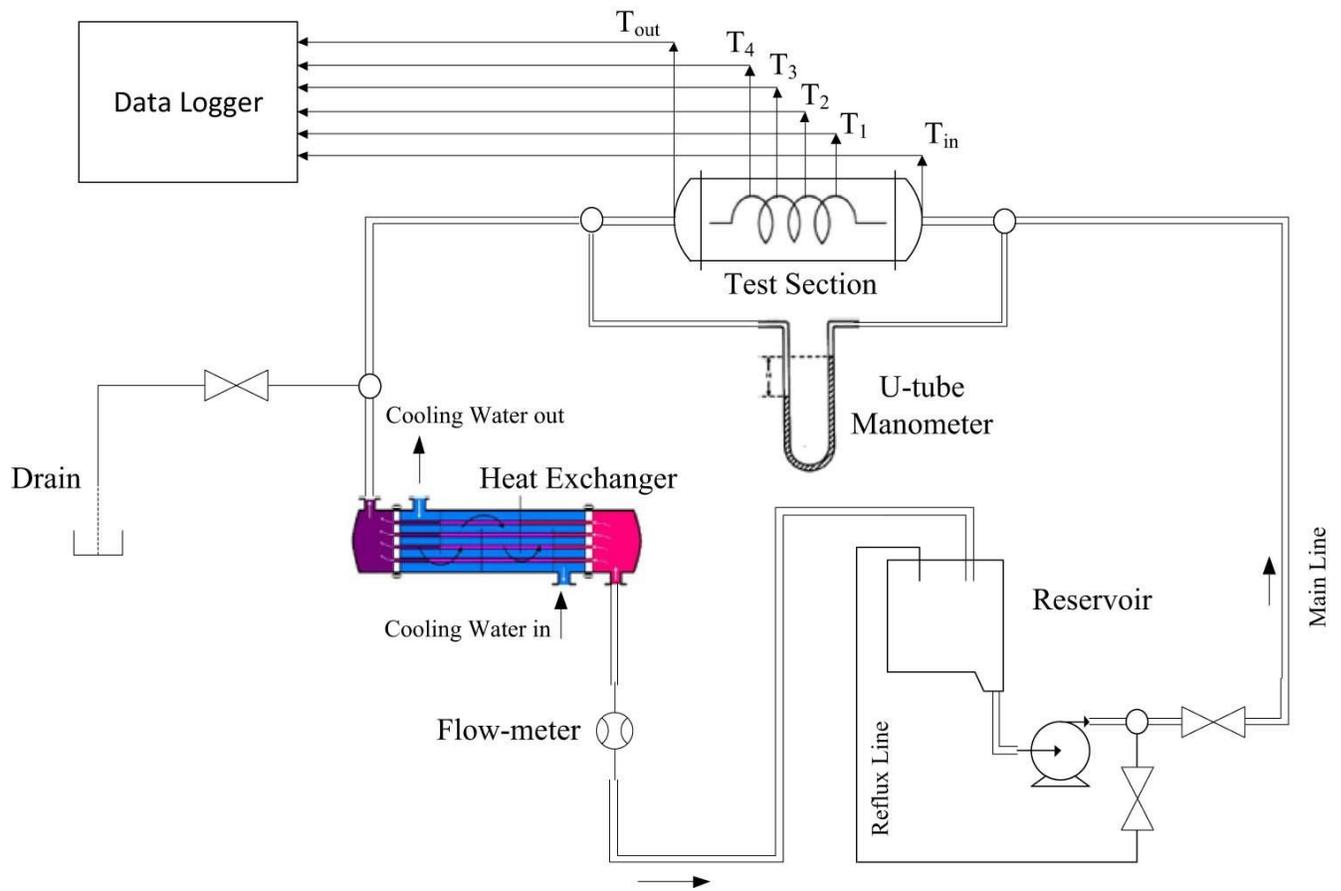

Figure 12. Experimental set up diagram ref [78].

Figures Figure 13 (a)-(c) show how the friction factor varies along helically coiled pipes for different Dean numbers at assorted curvature ratios. Typically, increasing Re lowers the friction factor due to changes in the velocity profile. However, for helically coiled pipes the Dean number primarily governs the friction factor trend, as depicted. Figures Figure 13 and Figure 14 demonstrate that at a constant curvature ratio, higher Dean numbers raise the friction factor. Physically, stronger secondary flows develop at greater Dean numbers, intensifying momentum transfer within the coiled pipes. Furthermore, enhanced secondary flows shift the maximum velocity closer to the wall, steepening the wall-normal velocity gradient and increasing friction. Therefore, the Dean number emerges as a key parameter influencing friction characteristics in helical pipe configurations.

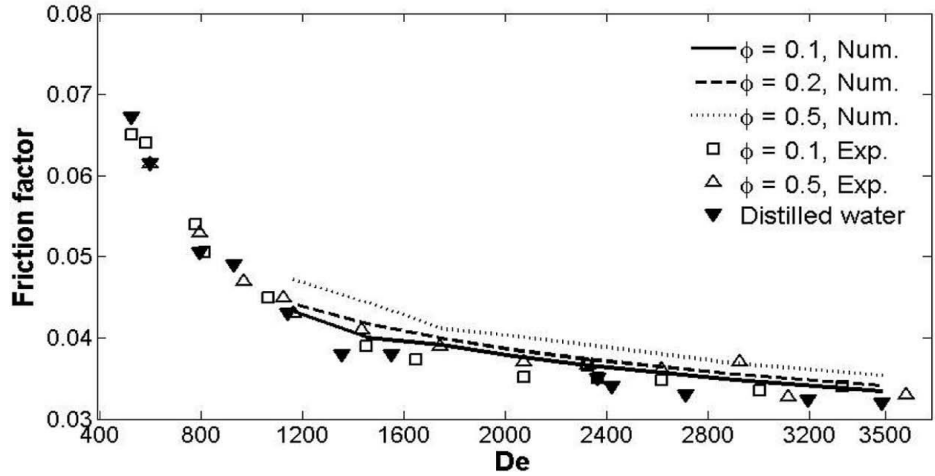

(a)

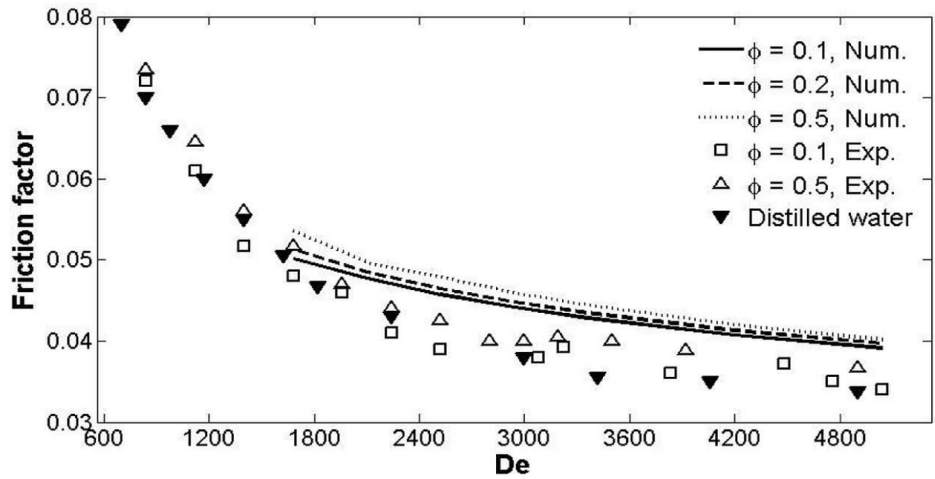

(b)

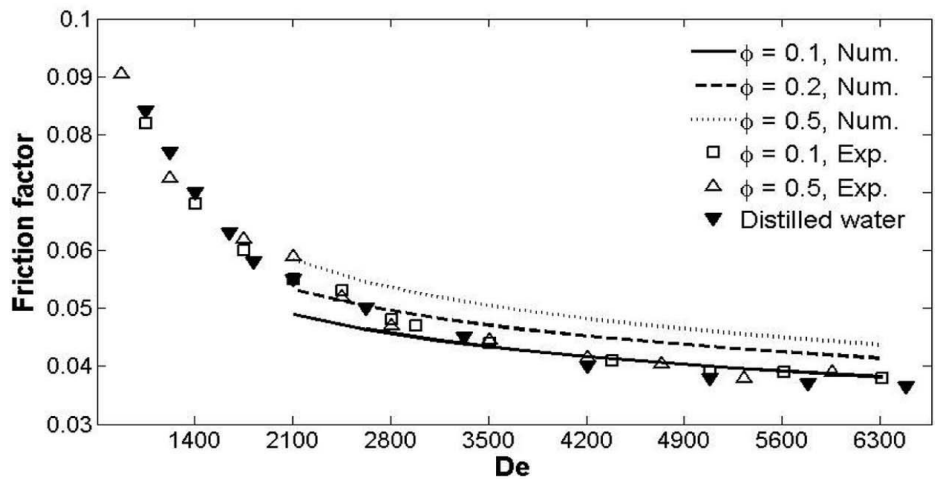

(c)

Figure 13. The relation between friction factor and dean number for (a) $\kappa = 0.0375$ (coil 1), (b) $\kappa = 0.0784$ (coil 4), and (c) $\kappa = 0.1231$ (coil 5) [78].

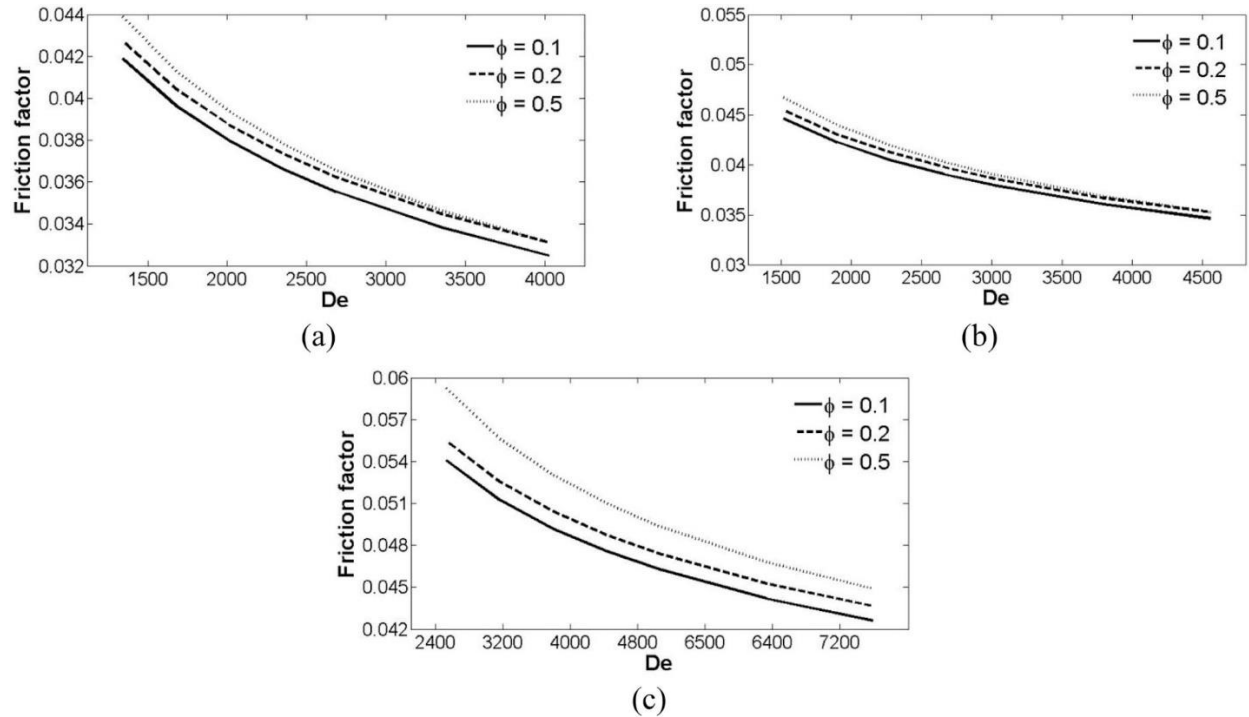

Figure 14. Variation of friction factor along the helically coiled pipe for different values of Dean number and for (a) $\kappa = 0.05$ (coil 2), (b) $\kappa = 0.064$ (coil 3), and (c) $\kappa = 0.1778$ (coil 6) [78].

Huu-Quan et al. [79] conducted CFD simulations to gain a clearer understanding of turbulent forced convection characteristics in novel double-pipe heat exchanger featuring flat inner pipe. Effects of variations in inner pipe geometry, specifically aspect ratio, on the flow behavior and heat transfer were numerically analyzed. The results indicated that at lower inner pipe Reynolds numbers below 7000, configurations with flat inner pipes possessing a small aspect ratio of 0.37 provided benefits over traditional circular pipe designs. Specifically, the flat inner pipe geometry led to approximately 2.9% higher overall heat transfer coefficient, 2.7% increase in thermal effectiveness, and a 16.8% boost to performance index. However, at higher Reynolds numbers exceeding 7000, circular inner pipes were found to significantly outperform the flat pipe options. Therefore, the study concluded that for double-pipe heat exchangers operating at lower flow velocities, adopting flat inner pipe with a compact aspect ratio can enhance performance, while circular pipes remain superior at higher flow rates.

Rakhsha et al. [80] numerically and experimentally investigated the flowing of CuO nanofluid inside helical tubes under turbulent forced convection at a constat wall temperature condition. In experiments, a heat exchanger with variable coiled tube geometries maintained a set wall temperature. Pressure drops and temperature measurements at the inlet and outlet were used to determine water and nanofluid Nu and friction factor. Numerical simulations employed the OpenFOAM solver to simulate the model. Transport properties were assumed constant based on the bulk mean temperature. Results showed an increase up to 6-7% in heat transfer and 9-10% higher pressure drop for the nanofluid versus water numerically. Experiments reported 16-17% enhanced heat transfer coefficients and 14-16% rise in pressure drop for different configurations and Reynolds numbers. Both pressure drops and heat transfer increased with greater curvature ratios and Re. Correlating equations were introduced to evaluate the friction factor and Nusselt number based on numerical and experimental findings. This comprehensive study explored fluid flow and heat transfer resulting from CuO nanoparticles in coiled tubes, gaining insights applicable to efficient heat exchanger tailoring.

Bhattacharyya et al. [81] conducted CFD simulations utilizing the transition SST turbulence model supplemented experimental analysis of forced convection in twisted tubes. This allowed investigating flow regimes ranging from laminar to transitional to turbulent as the Reynolds number ranged between 100 and 50,000. Simulations were conducted for a range of tube length and pitch ratios. The predictions showed good agreement with collected measurements and literature correlations. Analysis found that longer length ratios and smaller pitch ratios enhanced heat transfer performance at a relatively low-pressure penalty. Notably, the twisting tube geometry generated performance factor values exceeding unity for nearly all test conditions. This indicates the tube twisting configuration favorably augmented heat transfer over a straight tube design.

Wu et al. [82] numerically investigated turbulent forced convection in 3D twisted elliptical tubes with a constant surface temperature using CFD. They analyzed the effects of Reynolds number and twist pitch on flow resistance, heat transfer, and overall thermal-hydraulic performance of water flow. Results showed twisting motion in elliptical tubes enhanced synergy between velocity vectors and temperature gradients, improving heat transfer versus equivalent oval tubes. While the twisting geometry increased the encountered pressure drop due to disrupted flow, augmenting heat transfer performance. Specifically, a tube with a diameter of 96mm demonstrated a 58-60% rise in pressure drop but a 16-19% higher average Nusselt number relative to an oval configuration. Furthermore, variations in Reynolds number and twist pitch influenced Nusselt number and pressure drop trends, in addition to impacting the overall thermal performance. A twist pitch and tube diameter of 128mm exhibited the optimal balance.

### 3.3. Summary of turbulent correlations

The correlations of turbulent forced heat transfer convection have been investigated for decades externally, numerically, and theoretically for different tubes. These correlations could be for Nussle number, Reynolds number, and friction factor. A lot of studies investigated these correlations with nanofluid materials inside tubes for different heat exchangers and applications. Table 2 summarizes several turbulent correlations mentioned in the open literature, the table describe the tube geometry, correlation, study approach and some remarks.

Table 2. Summary of turbulent forced heat transfer convection correlations inside tubes reported in open literature.

| Tube types | Correlations | Remarks | Study approach | Ref. |
|---|---|---|---|---|
| Vertical 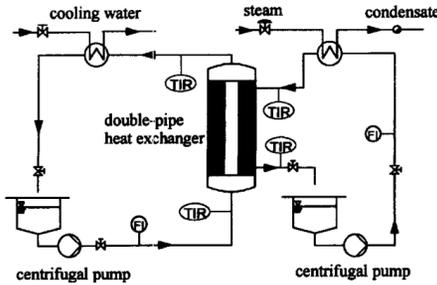 | For $Re \geq 10000$ $$Nu_{FT} = \frac{\frac{\xi}{8} \cdot Re \cdot Pr}{1+12.7 \cdot \sqrt{\xi/8} \cdot (Pr^{2/3}-1)} \left(1 + \left(\frac{d}{L}\right)^{\frac{2}{3}}\right)$$ $$\xi = (1.8 \cdot \log(Re) - 1.5)^{-2}$$ For $2300 < Re < 10000$ $$Nu = (1-\gamma) \cdot Nu_{FL}(Re = 2300) + \gamma \cdot Nu_{FT}(Re = 10000)$$ $$\gamma = \frac{Re - 2300}{10000 - 2300}$$ | The study examined the effects of varying length-to-diameter ratios and opposing or concurrent heat and mass flux directions on heat transfer within vertically oriented tubes. | Exp. | [49] |
| Helical tube 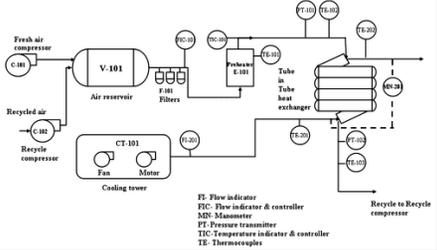 | $$Nu = 0.552\, Dn^{0.637} Pr^{0.4}$$ $1.4 \times 10^4 < Re < 8.6 \times 10^4$ $Pr = 0.70$ $\delta = 0.033$ | Water is used to cool the air inside the helical tube. | Experimental | [50] |
| Helical tube | $$Nu = 0.013\, Re^{0.93} \delta^{0.177} Pr^{0.4}$$ $9.7 \times 10^3 < Re < 1.4 \times 10^5$ $0.70 < Pr < 6$ $0.012 < \delta < 0.177$ | For Re larger than $10^5$, the effect of roughness in friction factor should be considered | Theoretical | |
| Helical tube 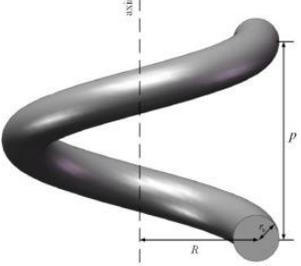 | $\dfrac{1}{\sqrt{f}} = -0.923 \ln \left( \dfrac{0.104\,\varepsilon\,\delta^{0.5}}{f} + \dfrac{1.142\,\delta^{0.5}}{Re\,f^{3/2}} \right)$<br><br>$1 \times 10^4 < Re < 1 \times 10^5$<br>$0.05 < \delta < 0.062$<br>$9.6 \times 10^{-5} < \varepsilon < 3.2 \times 10^{-4}$ | The study discussed the friction factor in a rough helical tube. The results indicated the criticality of accounting for roughness increases substantially with both curvature ratio and Reynolds number in helical tubes. | Theoretical and experimental | [57] |
| Circular tube 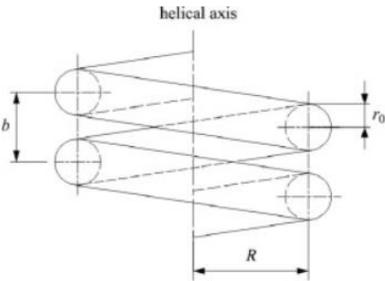 | $Nu = 0.5277 \cdot 10^{-2} Re + 0.1625 \cdot 10^5 \varphi_v^2$<br>$\quad + 0.3066 \cdot 10^3 \varphi_v + 0.3084 \cdot 10^2$<br>$1 \times 10^4 < Re < 1 \times 10^5$<br>$0.001 < \varphi_v < 0.04$ | They used Al$_2$O$_3$–water nanofluid with a variable tube. | Numerical | [83] |
| Horizontal circular tube | $Nu = 0.00218 Re^{1.0037}\ Pr^{0.5}\,(1 + \varphi_v)^{154.6471}$<br>$4800 < Re < 30500$<br>$0 < \varphi_v[\%] < 0.25$<br>$5.5 < Pr < 5.59$ | The study used TiO$_2$ Nanofluid inside tube, and investigated single phase and two phase models. | CFD | [84] |
| Circular tube | $Nu = 0.085\ Re^{0.71} Pr_r^{0.35}$<br>$10^4 < Re < 5 \times 10^5$<br>$0 \le \varphi_v[\%] \le 10$<br>$6.6 < Pr < 13.9$ | The study analyzed turbulent flow and convection heat | Numerical | [85] |
| | | transfer of a nanofluid, specifically a mixture of water and gallium oxide nanoparticles, inside a tube with a uniform wall temperature boundary condition. | | |
| Plain tube | $Nu_{nf} = 0.27\, Re^{0.693} Pr^{-0.3} \left(1 + \dfrac{D}{H}\right)^{1.3}$ | The correlation developed in the study is applicable for predicting heat transfer of a titanium dioxide-water nanofluid with nanoparticle concentrations up to 3.0%. | Experimental | [67] |
| Helical tube | $Nu = 0.028 Re^{0.854} Pr^{0.38} k^{0.039} \Phi^{0.041}$<br>And the fraction factor<br>$f = 0.302\, Re^{-0.177}\, k^{0.116} \Phi^{0.013}$<br><br>$2346 < Re < 18563$<br>$2.7 < Pr < 3.51$<br>$0.0375 < k < 0.123$<br>$10^{-3} < \Phi < 5 \times 10^{-3}$ | Constant wall temperature | Experimental and numerical | [78] |

| Horizontal straight tube | $Nu = 0.021 Re^{0.8} Pr^{0.5}$ | | Experimental | [86] |
|---|---|---|---|---|
| Helical tube 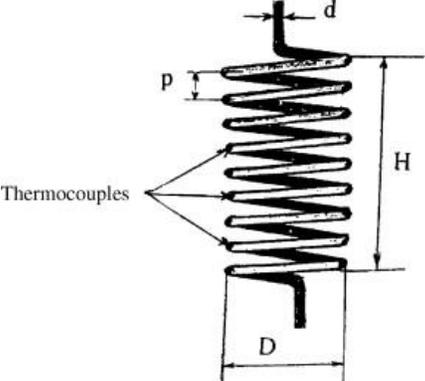 | $Nu = 0.0345 Re^{0.48} \left(\dfrac{D}{d_0}\right)^{0.914} \left(\dfrac{P}{d_0}\right)^{0.281}$ $7.086 \leq Re \leq 16.142$ $6.6 \times 10^2 \leq \dfrac{D}{d_0} \leq 2.3 \times 10^3$ $1.81 \leq \dfrac{P}{d_0} \leq 3.205$ | The study numerically examined forced convective heat transfer from the outer surfaces of helically coiled tubes that were subjected to a uniform heat flux boundary condition. | Experimental | [71] |

# 4. Conclusion and remarks

This review synthesized a comprehensive overview of forced convective heat transfer tube flow knowledge while also acknowledging ongoing needs such as developing generalized correlations spanning broader operating ranges. Continued focus on laminar and turbulent flows as well as tube shape impacts would further enrich available design tools. Overall, better understanding flow turbulence, boundary layers and materials could help optimize thermal systems critical to engineering applications. By thoroughly evaluating the extensive research basis, this study aimed to consolidate key forced convection learning applicable across disciplines. Further advancing the presented conclusions will contribute to progressing heat transfer science and its diverse industrial applications. We can summarize the following findings:

- The literature on experimental studies of heat transfer of nanofluids flowing inside tubes is quite extensive.
- Helical tubes are commonly examined in industrial applications, however further research is still required.
- Most numerical investigations have used the standard k-ε turbulence model to study nanofluid flows yet more work is needed to advance characterization of their thermophysical properties which heavily impact model accuracy.
- The limited forced heat transfer correlations proposed so far exhibit narrow applicability ranges without a generalized friction factor correlation derived for horizontal tubes.
- Additional research should develop comprehensive hydrodynamic and heat transfer correlations validated over broader conditions to better understand nanofluid behavior.
- Further studies are also warranted to investigate tube geometry influences like cross-sectional shape on flow characteristics within tubes and heat exchangers.
- Preparing and testing hybrid nanofluids composed of solid particles suspended in mixed base fluids may help optimize heat transfer for automotive cooling applications.
- It is challenging to experimentally achieve fully developed forced convection conditions with a constant heat flux boundary in laminar and transitional flow regimes due to difficulties like: Tests requiring tubes to have very small diameters and small heat fluxes to minimize buoyancy effects. This leads to high uncertainties from small temperature changes between surface and fluid. Despite difficulties, obtaining accurate experimental data with low uncertainties is important for fundamentally understanding internal forced convection heat transfer.